\documentclass[a4paper]{jpconf}

\usepackage{graphicx}

\begin{document}

\title{Glimpsing Colour in a World of Black and White}
\author{M.~R.~Pennington}
\address{Theory Center, Jefferson Laboratory, 12000 Jefferson Avenue, Newport News, VA 23606, U.S.A.}
\ead{michaelp@jlab.org}

\begin{abstract}
The past 40 years have taught us that nucleons are built of constituents that carry colour charges with interactions governed by Quantum Chromodynamics (QCD). How experiments (past, present and future) at Jefferson Lab probe colourless nuclei to map out these internal colour degrees of freedom is presented. When combined with theoretical calculations, these will paint a picture of how the confinement of quarks and gluons, and the structure of the QCD vacuum, determine the properties of all  (light) strongly interacting states.  
\end{abstract}

A hundred years ago, Rutherford demonstrated that the experimental results of his 
colleagues, Geiger and Marsden, pointed to there being  a tiny nucleus at the 
heart of every atom. This started off the field we celebrate at this meeting. For awhile it was believed that each nucleus was a collection of protons and neutrons held together by some strong nuclear force. However, over the past 50 years, we have learnt that inside nuclei not only are
protons and neutrons present, but also dozens, even hundreds of other short lived particles, 
we call collectively hadrons. Hadrons fall into two classes: baryons, of which the proton is the only stable 
example, are fermions, and mesons, of  which the pion is by far the lightest, are bosons. These could be understood if each were themselves built of constituents, quarks: each meson made of a quark and an antiquark, and every 
baryon with three quarks. In 1970 we knew of just three flavours of quark, up, 
down and strange, but in quick succession heavier quarks, charm and bottom, and 
then top were discovered. For this talk we restrict discussion to the world of just the 
lightest three flavours of quark, $u$, $d$ and $s$ of which we are made. This places the 
proton and neutron in an octet of ground state baryons, and the three differently 
charged pions in a nonet of ground state pseudoscalar mesons. Deeper probing, in the spirit of Rutherford, has taught us that there really are quarks inside hadrons, but however hard we hit them they never leave the femto-universe alone. Quarks can then not just be \lq\lq bits'' of  a proton or a neutron, but 
rather must have a property that nucleons don't have: a property we call colour. 
While 
hadrons are colour neutral, quarks carry colour bound by a force governed by Quantum Chromodynamics (QCD). 

This talk is a brief review of the work of Jefferson Laboratory (JLab) that sets out to understand these colour degrees of freedom.
JLab studies protons and neutrons in fine detail, alone and within nuclei, using an electron beam as a probe.
This includes important parity violation experiments that measure the interference between the photon and the $Z$-boson to study the weak charges of quarks and leptons. These experiments are covered in the talk by 
Michael Ramsey-Musolf~\cite{MRM}. Other parity violation experiments, like PREX,  use the weak interaction too to measure the neutron skin of nuclei, like 
lead~\cite{PREX}. 
Understanding this is important for the workings of astrophysical objects, like neutron 
stars. Here I will concentrate on hadron physics and the strong interaction at JLab.

\begin{figure}[h]
\begin{center}
\includegraphics[width=12.5cm]{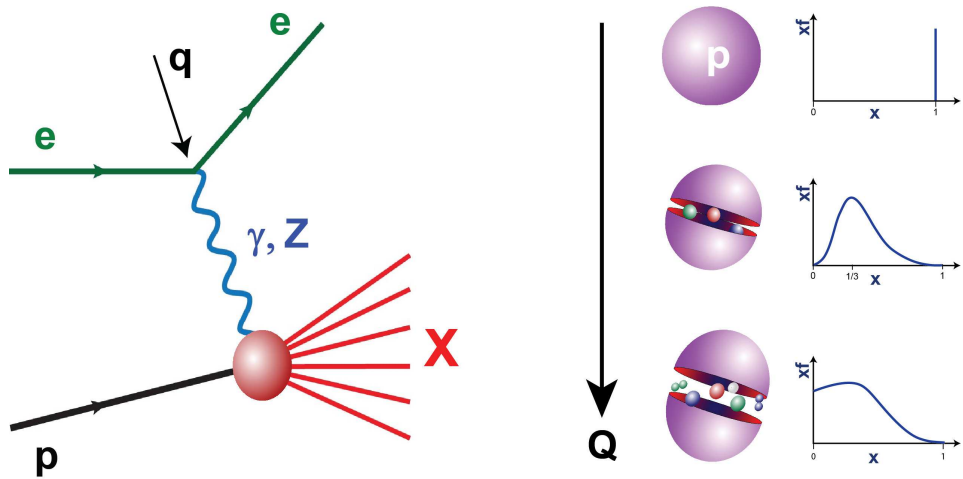}
\end{center}
\caption{Deep inelastic electron scattering on a proton is shown on the left. As the photon virtuality, $Q$ where $Q^2=-q^2$, increases (depicted on the right), the photon wavelength shortens and it probes the proton's internal structure at shorter scales. The cross-section can be expressed in terms of $xf(x)$ the momentum distribution  of the constituents of the proton. Bjorken $x$ is the fraction of its longitudinal momentum carried by the struck \lq\lq parton'' in the light cone or Breit frame. At low $Q$ (or long wavelength)  the photon sees the whole proton. As $Q$ increases, it begins to see the valence quarks, and then as the wavelength shortens further, the current quarks, sea of ${\overline q}q$ pairs and cloud of gluons. Above a $Q$ of $\sim 3$ GeV, the evolution of the momentum distribution $xf(x)$ sketched here is governed by perturbative QCD.}
\end{figure}

As the CEBAF accelerated electron approaches a nuclear target, it feels the proton charge and emits a virtual 
photon that probes inside the proton, Fig.~1. In a frame in which the nucleon moves at 
relativistic speeds, the photon sees the proton's longitudinal momentum shared by its 
valence quarks and a cloud of gluons. At higher energies, the 
photon having still shorter wavelength sees an increasingly important sea of quark and 
antiquark pairs each carrying a smaller and smaller fraction of the proton's 
momentum. This evolution is governed by perturbative QCD and is well understood. 
What JLab studies is what happens when we just open the nucleon box and peer 
inside, Fig.~1.   JLab experiments also investigate how these momentum distributions change 
when the nucleon is inside a nucleus. Then in principal the quarks no longer know 
which nucleon they belong to and move from one to the other. 

Of course, the 
components of the proton not only carry its longitudinal momentum, its flavour and spin, but are 
distributed in transverse space, Fig.~2. A new era of 
tomography of the proton has started~\cite{tomography}. HERMES, COMPASS and JLab@6 GeV~\cite{makins} have 
shown that generalized parton distributions (GPDs) and transverse momentum distributions 
can be studied in semi-inclusive deep inelastic processes and deeply virtual Compton 
scattering.  The extraction of these will require close cooperation between 
experimentalists and theorists: a feature of JLab physics particularly in the upcoming 
era of precision data. A major effort  is on the way to predict what QCD can tell us about 
this internal landscape, 
Fig.~2, as well as  interpret data in terms of QCD.

\begin{figure}[t]
\begin{center}
\includegraphics[width=12.8cm]{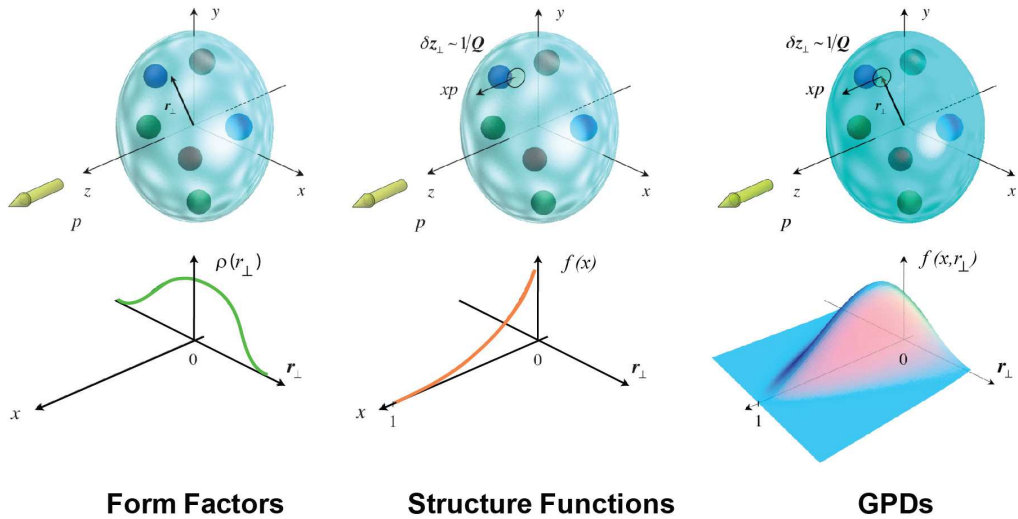}
\caption{With the hadron moving in the $z$-direction, the figure~\cite{tomography} illustrates how different physical quantities map both the longitudinal momentum ({\it i.e.} $x$) distribution and the transverse spatial distribution of the \lq\lq partons'' inside the hadron. Elastic form-factors involve just the transverse distribution, the structure functions of Fig.~1, the longitudinal momentum distribution, while generalized parton distributions (GPDs) depend on both variables.}
\end{center}
\vspace{-2.mm}
\end{figure}

This joint engagement of theory and experiment is also a key part of the program in  hadron spectroscopy, to which we now turn. With quark spins combined with units of orbital angular momentum we expect whole towers of 
hadrons beyond the ground states. For many, seeking these states is mere stamp-collecting. Indeed it was Rutherford who said \lq\lq all of science is either physics or 
stamp-collecting''. What I want to convince you of is that spectroscopy is physics. 
The paradigm is the spectrum of photons emitted by excited atoms. 
Even if we did not have enough energy to separate a nucleus from its surrounding 
electron cloud, the spectrum alone would teach us that atoms, though electrically 
neutral, behaved as if they were made of electrically charged objects held together by 
an electromagnetic force, governed by the rules of quantum electrodynamics. In an 
analogous way, colour neutral hadrons can teach us about the rules that lead to the 
confinement of quarks and thereby all the properties of light hadrons.  We believe we know 
the  underlying theory, namely quantum chromodynamics (QCD). At short distances, the {\it up} and {\it down} 
quarks are almost massless and interact only weakly with the octet of coloured gluons. 
Hard probing confirms QCD is correct to high 
precision in the short distance regime ($\ll 1 fm$). However, the properties of hadrons made of light quarks and glue are 
controlled by interactions at the distance of $\sim 1 fm$, where the strong interaction is 
strong and the vacuum filled with condensates. There accurate wholly theoretical calculations are much more difficult. Consequently, experiment provides critical guidance, which when combined with theory, is the key to understanding the colour degrees of 
freedom. 

While the colour neutral wavefunction of a meson is a sum  of quark-antiquark terms
regardless of the number of colours, that for a proton at its simplest involves $N_c$ 
valence quarks in $N_c$ colours, carrying electric charges in units of $1/N_c$. Baryons directly reflect the non-Abelian nature of 
QCD. But are hadrons just a collection of a minimal number of valence quarks?
Is the sole role of gluons to create the QCD vacuum that dresses almost massless $u$ and $d$ current quarks into constituent quarks of 300 MeV? Or can gluons form massive constituents themselves?

The motivation for the quark model in the first place was that it reproduced the low mass baryons found in experiment~\cite{dalitz}. The same model with three pointlike quark degrees of freedom~\cite{capstick} has long predicted
 the spectrum of excited nucleons and $\Delta$'s. 
However, experimentally
the excited spectrum above 1.6 GeV (or so) is far from clear. At first it was 
thought this was because the main source of information came from $\pi N$ elastic 
scattering, and perhaps the higher mass states coupled pre-dominantly to $\pi \pi N$ 
and $KY$ channels. More recent exploration of these final states in both hadro and electroproduction has revealed little more certainty from the gloom~\cite{pdg}. This is 
in part because multi-channel analyses are difficult, lengthy  and time-consuming. 
Nevertheless such analyses point to large numbers of expected excited baryons being absent. This could readily be explained~\cite{diquarks}, if the degrees 
of freedom in a baryon are restricted to a quark and a point diquark (with two of the valence quarks always pairing up). However appealing, is this modelling really QCD?

Moreover, in the 
simple quark model the spectrum computed is of stable states. The coupling  to $\pi N$, $KY$ and even $\rho N$ are popularly computed in a $^3P_0$ 
model of ${\overline q}q$ creation~\cite{barnes}. This allows an estimate of widths to  
decay channels, but how these feed back on the \lq\lq bare'' masses  is generally 
disregarded. Of course, heavier states are broader, spend less time in  their simplest 
Fock state but rather linger in multi-hadron modes, like $\pi\pi N$. These effects are not 
minor perturbations. 

\begin{figure}[b]
\begin{center}
\includegraphics[width=12.cm]{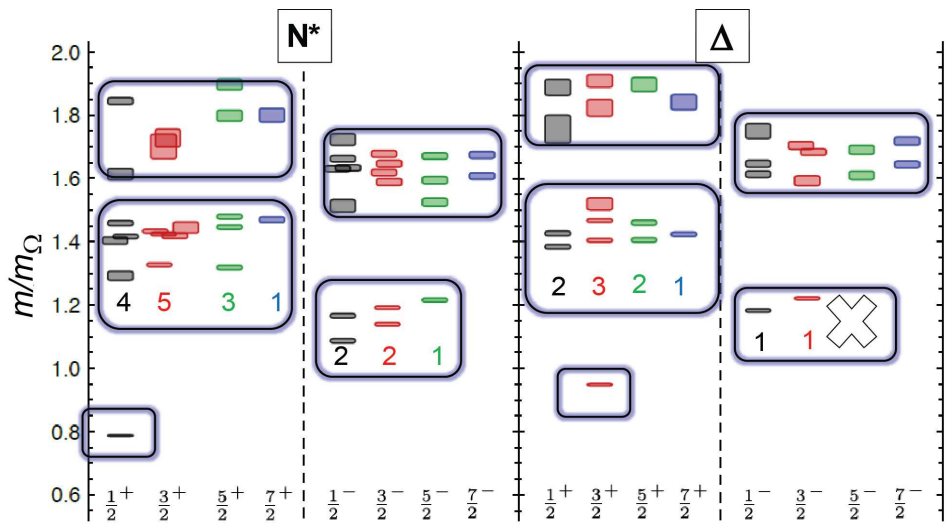}
\end{center}
\caption{Lattice calculation in full QCD of the baryon spectrum at a pion mass of 520 MeV from~\cite{edwards} labelled by $J^P$ in units of the $\Omega$ mass, showing the groupings into $SU(6)\times O(3)$ multiplets. }  
\end{figure}
Strong coupling QCD has long been studied on the lattice. Nevertheless, it is only now~\cite{edwards} 
that the baryon spectrum has been computed in fine detail with sufficient precision to 
reveal that the calculated states are naturally grouped in the $SU(6)\times O(3)$ 
structures of the 3-valence quark model, Fig.~3, with no support for pointlike diquark 
components. The experimentally missing states should be there. While hadronic 
channels are implicit among the operators used in these lattice calculations, the pion 
mass is 5 (and at best 3) times its physical value. Consequently decay channels have 
limited effect. This is an issue  the lattice must urgently address if it is to 
relate to the experimental light hadron sector.

\begin{figure}[t]
\begin{center}
\includegraphics[width=9.7cm]{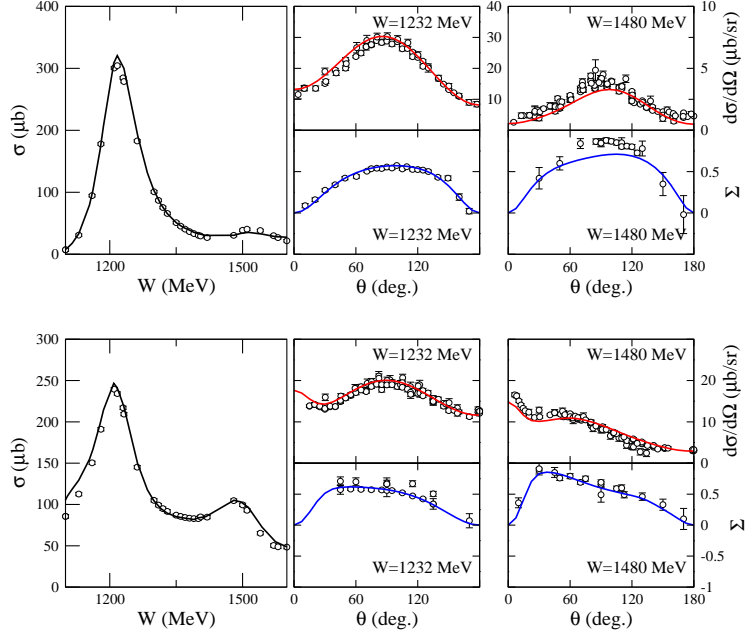}
\vspace{-2mm}
\caption{The upper 5 graphs are for $\gamma p\to\pi^0 p$ and the lower 5 for $\gamma p \to \pi^+ n$. On the left are the integrated cross-sections as functions of c.m. energy, $W$. On the right are differential cross-section and polarisation asymmetry, $\Sigma$, as functions of the c.m. scattering angle $\theta$, from~\cite{kamano}. }
\end{center}
\vspace{-5mm}
\end{figure}

\begin{figure}[b]
\begin{center}
\includegraphics[width=10.5cm]{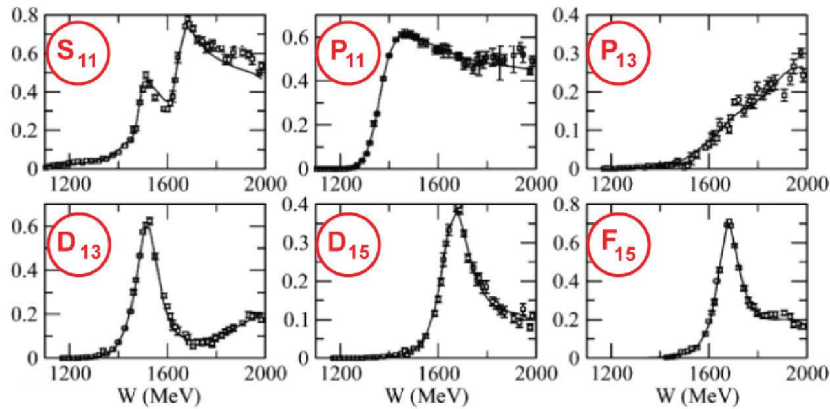}
\vspace{-2mm}
\caption{The imaginary parts of the lowest $\pi N$ partial waves are shown as a function of centre of mass energy, $W$ from the EBAC analysis of Julia-Diaz {\it et al.}~\cite{julia}. The waves are labelled by the orbital angular momentum $L$ of the $\pi N$ system, its isospin $I$ and total spin $J$ as $L_{2I\, 2J}$. As an illustration only those with $I=1/2$ are shown. }
\end{center}
\vspace{-5mm}
\end{figure}
Experience from heavy flavour factories has revealed an unexpected richness of the 
charmonium spectrum above $D{\overline D}$ threshold. While below, the spectrum 
is essentially that of a non-relativisitic potential, above open charm thresholds, decay 
channels not just distort the spectrum but add new states that are generated by 
inter-hadron forces. Some of these are remarkably narrow. Such dynamics is highly 
specific to the proximity of nearly open and nearby virtual hadron channels.
In other configurations, 
calculations show that states in the spectrum without decays can just as 
likely merge into the continuum, when open channels are included.

Similarly for baryons, coupled channels are essential to all states above the nucleon. Such hadron modes are an integral part of the mammoth EBAC effort~\cite{lee}. There within a 
specific Lagrangian framework a comprehensive list of meson, photon and electron 
induced baryon production channels are fitted, differential cross-sections and polarisation asymmetries, for instance those in Fig.~4 on $\gamma N \to \pi N$~\cite{kamano}, starting from the results of the $\pi N$ 
phase shift analysis of the SAID group~\cite{said}. 
 On the basis of such fits to some $10^5$ datapoints, one can determine the mass, width and couplings of excited baryon resonances. The resulting lowest $I=1/2$ $\pi N$ waves~\cite{julia} are illustrated in Fig.~5, as an example. Moreover, within such a modelling as that of EBAC~\cite{ebacroper}, hadron channels can 
be tuned in and out. A striking effect is in the $P_{11}$ channel, where the physical 
states, the $N^*(1440)$ (the Roper), and the $N^*(1710)$ reside as poles in the 
complex energy plane. When $\eta N$ and $\pi\Delta$ decay channels are switched 
off, these two poles move to the real axis merging as a single entity at $\sim 1800$ 
MeV. Perhaps it is this  single \lq\lq bare'' state that should be compared to the simple quark 
model, or even dynamical calculations in the Schwinger-Dyson/Bethe-Salpeter Equation approach~\cite{roberts}, and not two $N^*$'s.

However, the rules of the dynamics, viz QCD, that bind these quarks into hadrons 
lead us to expect an even richer spectrum, in which multiple ${\overline q}q$ pairs 
as well as gluons contribute to the quantum numbers of hadrons. As already alluded to, 
the charmonium sector gives strong hints of multiquark or molecular states, to which 
the strangeonium sector of the $f_0/a_0(980)$ can be added. However, finding states 
in which glue is an essential component, not just in generating binding as it does for 
all hadrons, but contributing as a basic degree of freedom would reveal much about 
the longer range dynamics  present inside all hadrons. It is in the meson sector that it 
is a little easier to predict how this might be present.
\begin{figure}[t]
\begin{center}
\includegraphics[width=5.5cm]{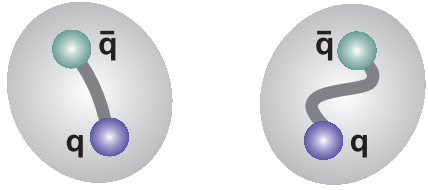}
\end{center}
\caption{Sketch of a typical quark-antiquark meson on the left, where the gluon flux just holds the constituent quarks together, while on the right is a {\it hybrid} meson, in which the gluon flux contributes in an essential way to the angular momentum of the hadron} 
\vspace{-2.5mm}
\end{figure}

The quark model of mesons gives simple rules determining the allowed spin $J$, parity $P$ and charge 
conjugation $C$, denoted collectively as $J^{PC}$. Finding mesons outside of the 
permitted values has been a thrust of meson spectroscopy for several decades. Such discovery would point to glue, or additional ${\overline q}q$ pairs, being essential to 
their make-up, Fig.~6. 
\begin{figure}[b]
\begin{center}
\includegraphics[width=10.3cm]{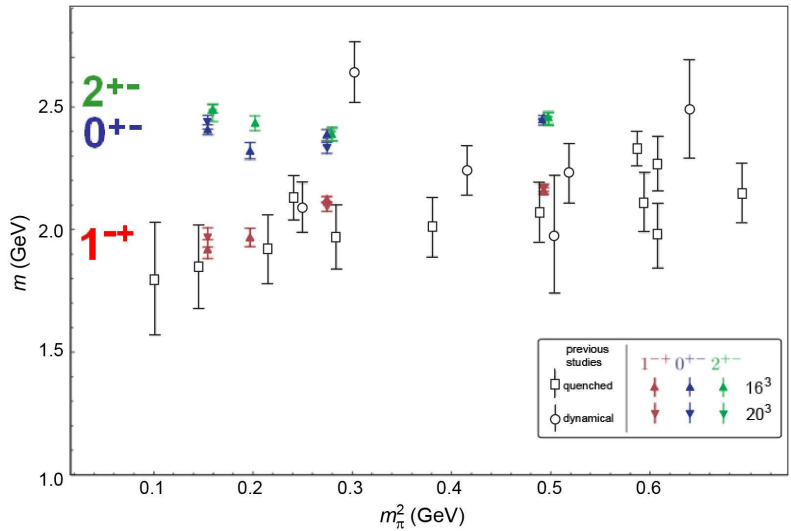}
\end{center}
\vspace{-6.mm}
\caption{Lattice QCD calculations of {\it hybrid} mesons with quantum numbers $J^{PC}\, =\, 1^{-+}$, $0^{+-}$ and $2^{+-}$ as a function of quark mass, expressed in terms of the square of the pion mass, $m_\pi$. The lattice data have to be extrapolated to the physical point, which is essentially where the values of $J^{PC}$ are drawn. Older lattice results are in black. The newer ones of higher precision are in full QCD from Dudek {\it et al.}~\cite{dudek}.}
\end{figure}
On the lattice, colleagues at JLab~\cite{dudek} have in full ({\it i.e.} unquenched) QCD computed the 
spectrum with ${\overline q}g q$ operators with unprecedented precision. These predict 
$1^{-+}$ states below 2 GeV, with $0^{+-}$ and $2^{+-}$ heavier in steps of 
several hundred MeV --- Fig.~7.

Experimental hints of a $1^{-+}$ enhancement started with GAMS results~\cite{gams} on $\pi^-p\to\pi^0\eta n$ at high 
energy and small momentum transfers. What was novel was that both the $\pi^0$ and 
$\eta$ were identified by their two photon decays. The $\pi\eta$ spectrum showed not only the 
well known $a_2(1230)$ and $a_0(980)$, but also revealed a forward backward 
asymmetry in the 1.4 GeV region that hinted at a $P$-wave effect with $1^{-+}$ 
quantum numbers, but this was never shown to be resonant. The VES experiment studied $\pi^- N\to \eta\pi^- N$ and $\eta'\pi^- N$ at 37 GeV/c~\cite{ves}. They similarly found an enhancement in the $\eta\pi$ amplitude around 1400 MeV (Fig.~8 left),  with some small phase change relative to the $2^{++}$ wave, and a broad enhancement $\sim 1600$ MeV in $\eta'\pi$.  
\begin{figure}[t]
\begin{center}
\includegraphics[width=13.cm]{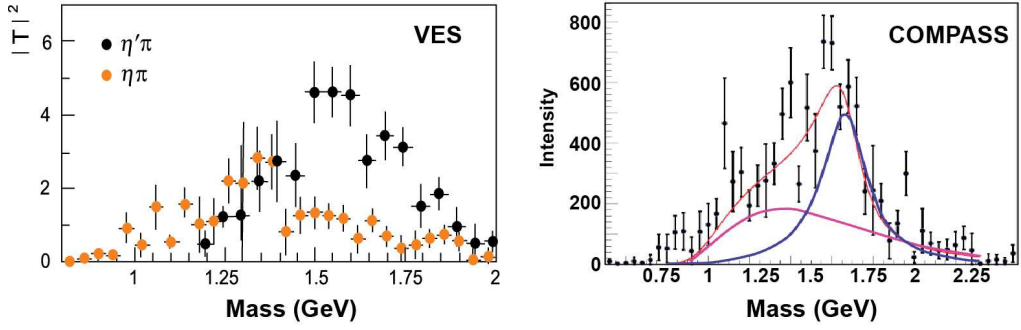}
\end{center}
\vspace{-5mm}
\caption{As functions of di-meson mass are shown with $J^{PC}\,=\,1^{-+}$: the modulus squared of the amplitude, $T$,  for the production of $\eta\pi$ and $\eta'\pi$ extracted from data on $\pi^- Be$ interactions at 37 GeV from VES~\cite{ves} on the left and for the $\rho\pi$ isobar (Fig.~9 left) decaying to $3\pi$  from the COMPASS $\pi^-Pb$ experiment~\cite{compass} at 190 GeV (on the right) --- fitted to a Breit-Wigner form.}  
\vspace{-2mm}
\end{figure}

\begin{figure}[b]
\begin{center}
\includegraphics[width=11.cm]{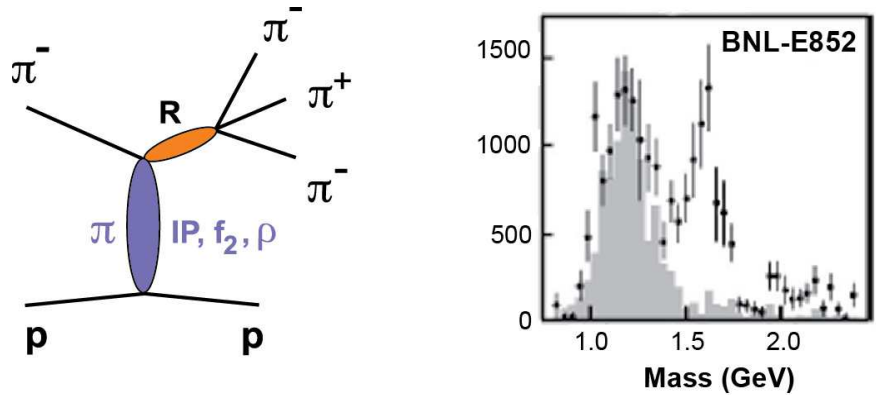}
\end{center}
\vspace{-3mm}
\caption{Regge exchange picture of three pion production in $\pi^-p$ interactions at small momentum transfers (on the left) assuming an isobar $R$ is first formed. On the right is the extracted $R\,=\,\rho\pi$ signal with $J^{PC}\,=\,1^{-+}$ from BNL-E852 data~\cite{chung2}. The grey histogram is the calculated \lq\lq leakage'' into this channel from other partial waves. The enhancement at $\sim 1.4$ GeV is thereby explained~\cite{dzierba}, but leaves a clean $\sim 1.6$ GeV enhancement, that might be resonant.}
\end{figure}
A higher statistics BNL 
experiment  studied both the $3\pi$ and $\eta'\pi$ final states. These showed what 
might be a resonant peak around 1.4 GeV at an intensity of a few percent~\cite{chung1}, Fig.~9 right. To be 
certain it is there, one must understand other waves to better than this accuracy. 
Indeed, subsequent analysis showed that leakage from other partial waves could 
entirely produce this effect~\cite{dzierba} (Fig.~9). However, this was not the case for a higher mass peak 
and a $\pi_1(1600)$ was claimed~\cite{pdg}.  However, its signal in the $3\pi$ channel could 
readily change by a factor of 4 depending on how the various dipion final state 
interactions were treated~\cite{chung2}. More recently,  COMPASS have seen a broad $1^{-+}$ 
signal in $3\pi$~\cite{compass}, Fig.~8 right. 
However, a clean and clear phase variation has not been 
observed. The analysis of multi-particle final states is complicated with often 20-30 
partial waves components required even within a simple isobar picture, Fig.9.

The 12 GeV upgrade at JLab will not only sharpen our view of the internal 
landscape of the nucleon, but also dramatically improve the image of the 
hadron spectrum. In particular, a new experimental hall D will house the GlueX 
detector specifically designed to investigate the photoproduction of multi-hadron final 
states into which exotic mesons and baryons may decay, specifically those that may 
reveal the role that gluons play.  This mission is also 
shared in part by the upgraded hall B detector CLAS12.  While experiments will not start till 
2014 or even '15, plans are afoot to bring together a worldwide team of theorists to 
work with these groups to set the framework for the complex analyses of final state 
interactions that is required with precision data to unambiguously reveal structures in  
partial waves, in both  moduli and phases, that can point to gluonic degrees of 
freedom. Such a framework is essential to make the most of the 
statistics already available with BaBar and Belle data, daily produced in BESIII and 
LHCb, and yet to come with PANDA@FAIR and JLab@12 GeV.

With signals that are probably only there at a few percent, one needs analyses of even 
better accuracy. For the most part, present techniques are suited to data an order of 
magnitude poorer. There is little point taking data of the planned precision without 
analysis techniques to match. In this way one hopes to be certain to discover hybrid mesons, if they exist, not just with one set of quantum numbers, but in whole flavour multiplets. Similarly, light will hopefully be shed on the \lq\lq dark'' baryon sector. This will usher in an era of certainty in  hadron spectroscopy from 1.5 to 2.5 GeV, that with parallel theoretical effort will teach us how QCD, through the confinement of colour and the structure of the vacuum, determines the properties of all the hadrons in the nuclear world. Rutherford started this field in black and white. Little could he have known that we would be colouring in the picture a hundred years later.

\ack{Thanks are due to the organisers, especially Sean Freeman and Dawn Stewart, for the invitation to talk at this historic meeting. The work was authored in part by Jefferson Science Associates, LLC under U.S. DOE Contract No. DE-AC05-06OR23177.}


\vspace{5mm}
\begin{thebibliography}{99}
\bibitem{PREX} Horowitz CJ, Pollock SJ, Souder PA and Michaels R 2001 {\it Phys.\ Rev.}\ C {\bf 63} 025501
\bibitem{MRM} Ramsey-Musolf M, these proceedings and references therein
\bibitem{tomography} Belitsky AV and Radyushkin AV 2005 {\it Phys.\ Rept.}\ {\bf 418} 1
\bibitem{makins} Makins NCR 2011{\it Proc PANIC11} Cambridge, MA, USA July 2011 (to appear)
\bibitem{dalitz} Dalitz RH 1966 {\it High Energy Physics Les Houches 1965} eds. de Witt C and Jacob M (New York: Gordon \& Breach) pp. 251-323
\bibitem{capstick} 
  Capstick S and Isgur N
  1986 {\it Phys.\ Rev.}\  D {\bf 34} 2809;
  Capstick S and Roberts W
  1994 {\it Phys.\ Rev.}\  D {\bf 49} 4570
\bibitem{pdg} Nakamura K {\it et al.} (PDG) 2010 {\it J.\ Phys.}\ G {\bf 37} 075021
\bibitem{diquarks} Lichtenberg DB and Tassie LJ 1967 {\it Phys.\ Rev.} {\bf 155} 1601
\bibitem{barnes} Micu L 1969 {\it Nucl.\ Phys.}\ {\bf B10}, 521; Le Yaouanc A, Olivier L, P\'ene O and Raynal J 1973 {\it Phys.\ Rev.}\ {\bf D8} 2223; Ackleh ES, Barnes T and Swanson ES, 1996 {\it Phys.\ Rev.}\ {\bf D54} 6811
\bibitem{edwards} Edwards RG, Dudek JJ, Richards DG, Wallace SJ, arXiv:1104.5152 [hep-ph]
\bibitem{lee} Matsuyama A, Sato T and Lee TS
  2007 {\it Phys.\ Rept.}\  {\bf 439} 193
  [arXiv:nucl-th/0608051]
\bibitem{kamano} Kamano H, Julia-Diaz B, Lee T-SH, Matsuyama A and Sato T  2009 {\it Phys.\ Rev.}\ C {\bf 80} 065203
\bibitem{said} Arndt RA, Briscoe WJ, Paris MW, Strakovsky II and Workman RL
 2009 {\it Chin.\ Phys.}\  C {\bf 33} 1063
  [arXiv:0906.3709 [nucl-th]]
\bibitem{julia} Julia-Diaz B, Lee T-SH, Matsuyama A and Sato T 2007 {\it Phys.\ Rev.} \ C {\bf 76} 065201 [arXiv:0704.1615 [nucl-th]]

\bibitem{ebacroper} Kamano H, Nakamura SX, Lee TS and Sato T
  2010 {\it Phys.\ Rev.}\  C {\bf 81}, 065207
  [arXiv:1001.5083 [nucl-th]]
\bibitem{roberts} see for instance: 
  Eichmann G, Cloet IC, Alkofer R, Krassnigg A and Roberts CD
  2009 {\it Phys.\ Rev.}\  C {\bf 79} 012202
  [arXiv:0810.1222 [nucl-th]
 \bibitem{dudek} Dudek JJ, Edwards RG, Peardon MJ, Richards DG and Thomas CE 2010 {\it Phys.\ Rev.}\ {\bf 82} 034508
\bibitem{gams}Alde  D {\it et al} (GAMS) 1988 {\it Phys. Lett.} {\bf 205B} 397
\bibitem{ves} Beladidze GM {\it et al.} (VES) 1993 {\it Phys.\ Lett.}\ B{\bf 313} 276
\bibitem{chung1}Thompson  D R {\it et al.} (BNL-E852), 1997 {\it Phys. Rev. Lett.}\ {\bf 79} 1630;
Adams  GS {\it et al.} 1998 {\it Phys. Rev. Lett.} {\bf 81} 5760

\bibitem{dzierba}
Dzierba  AR {\it et al.} 2003 {\it Phys. Rev.} {\bf D67}  094015

\bibitem{chung2} Chung SU {\it et al.} (BNL-E852) 1999 {\it Phys.\ Rev.}\ {\bf D60} 092001 [hep-ex/9902003]
\bibitem{compass} Alekseev MG {\it et al.} (COMPASS) 2010 {\it Phys.\ Rev.\ Lett.}\ {\bf 104} 241803 [arXiv:1001.4654[hep-ex]]
\end{thebibliography}
\end{document}